\documentclass[twocolumn,amsmath,amssymb, prd, superscript address]{revtex4}

\usepackage{graphicx}
\usepackage{dcolumn}
\usepackage{bm}
\usepackage{epstopdf}%

\begin{document}

\preprint{MITCTP . . .}

\title{General Issues Connecting Flavor Symmetry and Supersymmetry}

\author{Ersen Bilgin}
\affiliation{Department of Physics, Williams College, Williamstown, Massachusetts 01267, USA}
\author{Brian Patt}
\affiliation{Center for Theoretical Physics, Laboratory for Nuclear Science and Department of Physics, Massachusetts Institute of Technology, Cambridge, Massachusetts 02139, USA}
\author{David Tucker-Smith}
\affiliation{Department of Physics, Williams College, Williamstown, Massachusetts 01267, USA}
\author{Frank Wilczek}
 \affiliation{Center for Theoretical Physics, Laboratory for Nuclear Science and Department of Physics, Massachusetts Institute of Technology, Cambridge, Massachusetts 02139, USA}

\date{\today}

\begin{abstract}
We motivate and construct supersymmetric theories with continuous flavor symmetry, under which the electroweak Higgs doublets transform non-trivially.   Flavor symmetry is spontaneously broken at a large mass scale in a sector of gauge-singlet fields; the light Higgs multiplets naturally emerge as special linear combinations that avoid acquiring the generic large mass.  Couplings of the light Higgs doublets  to light moduli fields from the singlet sector could lead to important effects in the phenomenology of the Higgs sector.
\end{abstract}

\maketitle

\section{\label{sec:intro} Motivation and Introduction}

The standard model gives a principled and economical description of gauge interactions, but its accommodation of fermion masses and mixings is loosely constrained and profligate of parameters.    It seems quite unsatisfactory to regard these many parameters as coefficients in the fundamental equations of physics, which is how they appear in the standard model.  One could hope that they emerge from the solution of more compelling equations, yet to be discovered.    

For one of the parameters in question, namely the overall phase of the quark mass matrix, essentially the so-called theta parameter $\theta_{\rm QCD}$ of QCD, there is an attractive proposal of this kind, that is the Peccei-Quinn mechanism \cite{Peccei:1977hh}.   That mechanism can be implemented by augmenting the standard model to contain two Higgs doublets which carry appropriate charges under a continuous (anomalous) $U(1)$ symmetry.   It explains why $\theta_{\rm QCD} \lesssim 10^{-9}$ is found to be extremely small in nature, which otherwise seems quite mysterious and conspiratorial.  The Peccei-Quinn $U(1)$ symmetry modulates the overall phase of the quark mass matrix (the phase of its determinant, to be precise).  Having gone that far, it seems natural to ask whether additional aspects of fermion mass matrices might be modulated by (dynamically broken) symmetry operations.    

If we turn off the masses and mixing angles, by decoupling the Higgs field, the Lagrangian of the Standard Model exhibits a large flavor symmetry acting on the three generations of quarks and leptons.    We can maintain that symmetry, which of course contains and vastly extends the Peccei-Quinn symmetry, by promoting the usual Higgs field to a whole matrix of fields.   At first sight this appears to involve us in great difficulties, for the many Nambu-Goldstone bosons that arise will mediate exotic flavor-changing decays at unacceptable rates, as would the many weak-scale scalars; also, this set-up utterly trashes the successful unification of couplings \cite{Dimopoulos:1981zb}, which at present appears to provide our most convincing insight into physics beyond the standard model.   On reflection, however, we recall that similar difficulties already arose in the earliest concrete implementations of Peccei-Quinn symmetry.   In that context they were solved by postulating that that symmetry is spontaneously broken by (standard model) gauge singlets acquiring a vacuum expectation value well above the electroweak mass scale \cite{Kim:1979if}.    So we should try to generalize that feature as well.

In the generic case, if we break our extended symmetry at a large scale $f$, then the fields from our Higgs complex will be of two types: those that acquire masses of order $f$, and derivatively-coupled electroweak singlet Nambu-Goldstone fields.   Neither type can fulfill the usual role of the Higgs fields in the standard model.    The situation changes drastically in the context of supersymmetric models.   In supersymmetric models, it is not unusual for particular, generally complicated additional combinations of fields to remain massless, even though they are neither derivatively coupled nor singlets under symmetries that commute with the broken symmetry (flat directions).   Just this feature, in a broader context, makes supersymmetry an attractive hypothesis to address the disparity between electroweak and unification symmetry breaking scales.    Of course, the hypothesis of low-energy supersymmetry also plays an important quantitative role in insuring accurate coupling constant unification, which is an independent, powerful motivation to adopt it.   

Putting these considerations together, we are led to seek models in which the two Higgs doublets of the minimal supersymmetric extension of the standard model (MSSM) arise as ``survivors'' from a primary stage of flavor symmetry breaking at a very high mass scale.   (To preserve the unification of gauge couplings, it is safe as well as natrual to take the flavor symmetry breaking scale $f$, which determines the masses of the heavy Higgs doublets,  to be near or above the unification scale, $f \gtrsim 10^{16}$~GeV.)   The interactions of these Higgs fields with standard model fermions, that govern their masses and mixings, directly reflect the complicated conditions that pick out the survivor fields from among their massive brethren.  

The fields of the singlet flavor symmetry breaking sector likewise fall into three classes: massless Nambu-Goldstone bosons associated with the spontaneous breaking of the flavor symmetry, heavy singlets that acquire masses of order $f$ from the flavor symmetry breaking, and moduli fields, that are massless in the absence of supersymmetry breaking.   The Nambu-Goldstone fields will be derivatively coupled.  This implies that their couplings to standard model fields will involve interactions of mass dimension $>4$, and therefore will be suppressed by powers of $1/f$.   But since the light moduli fields are not derivatively coupled, they can piggyback on the only invariant field combination involving standard model fields with dimension $<4$, i.e.  $\phi^\dagger \phi$ for the Higgs doublet $\phi$, to couple unsuppressed.   It is also plausible, as we shall see, that flavor non-singlet moduli fields with weak-scale vacuum expectation values could play a role in addressing the $\mu$ problem, again leading to unsuppressed couplings.   Such couplings could have major implications for future collider phenomenology.  

Non-supersymmetric theories with Higgs flavor multiplets have been considered {\it e.g.} in \cite{Wilczek:1978xi}, and either have extra Higgs doublets at low energies, or  require fine-tuning
at tree level to achieve the desired mass separation between the light and heavy Higgs doublets.  
Existing supersymmetric models also feature extra light Higgs doublets \cite{Babu:2004tn}, unless a fortuitous arrangement of mass parameters is assumed \cite{Babu:2002ki}.  Our aim here is to develop reasonably simple hypotheses that accommodate realistic fermion masses and generate a single pair of light Higgs doublets in a robust manner.

For purposes of this paper we shall employ continuous global symmetries.  We are not unaware of theoretical objections to such symmetries, based on the hypothetical existence of gravitational corrections; but we see no reason why such corrections, if they exist at all,  could not be small quantitatively.  For further discussion, see \cite{Kallosh:1995}.

\section{\label{sec:su3} An $SU(3)$ Example}

 As a concrete illustration, we  take the flavor symmetry to be $SU(3)$, with the quark and multiplets all transforming in the fundamental representation: $Q$,  $U^c$, $D^c$, $L$, and $E^c \sim {\bf 3}$.  
We take the Higgs multiplets to be (${\bf 3}_u+{\bf {\overline 6}}_u+{\bf 8}_u$) and (${\bf 3}_d+{\bf {\overline 6}}_d+{\bf 8}_d$), where the fields with subscript $u$ (respectively $d$) carry hypercharge $+\frac{1}{2}$ ($-\frac{1}{2}$).  

The ${\bf 3}_u$,  ${\bf {\overline 6}}_u$, ${\bf 3}_d$, and  ${\bf {\overline 6}}_d$  multiplets all have renormalizable Yukawa couplings with the quarks and leptons.  The adjoints ${\bf 8}_u$ and ${\bf 8}_d$ do not, but play an important role in determining the final Higgs spectrum.  Suppose, in particular, that in the superpotential flavor-breaking fields  couple the flavor-adjoint Higgs fields to the ${\bf 3}$ and ${\bf {\overline 6}}$, but not the ${\bf 3}$ and ${\bf {\overline 6}}$ to each other.   Then at most eight independent linear combinations of the  nine up-type Higgs doublets contained in $ {\bf{\overline 6}}_u$ and ${\bf 3}_u$ can couple to the eight down-type Higgs doublets contained in ${\bf 8}_d$.  It follows that at least one up-type Higgs doublet remains massless; similarly, at least one down-type Higgs doublet remains massless.  For generic choices of the flavor-breaking fields, only the mandatory pair of Higgs doublets will remain massless.   

We can realize this scenario by taking the flavor-breaking sector to contain electroweak-singlet flavor multiplets ${\bf {6}}_s$ and ${\bf {\overline {3}}}_s$.   By  imposing a global $U(1)_X$ symmetry under which the charges of the quarks and leptons are given by the standard embedding  $SU(5) \otimes U(1)_X \subset SO(10)$, viz. $Q(1)$, $U^c(1)$, $D^c(-3)$, $L(-3)$, $E^c(1)$, $N(5)$, we can forbid many unwanted couplings.   The $U(1)_X$ charges of the multiplets that contain the light Higgs doublets are ${\bf 3}_u(-2)$, ${\bf {\overline 6}}_u(-2)$, ${\bf 3}_d(2)$, ${\bf {\overline 6}}_d(2)$.  For reasons that will emerge shortly, we take the charges for the flavor-breaking fields to be ${\bf {6}}_s(-5)$ and ${\bf {\overline {3}}}_s(-5)$.  Finally, to allow for the desired Higgs couplings in the superpotential, we choose the charges of the flavor-adjoints  to be ${\bf 8}_u(3)$ and ${\bf 8}_d(7)$.

The most general superpotential involving the flavor-breaking fields and Higgs doublets consistent with these $SU(3)$ and $U(1)_X$ symmetries is 
\begin{eqnarray}\label{eq:superpotential}
W & = & {\bf {\overline 6}}_u {\bf 8}_d {\bf {6}}_s +{\bf {\overline 6}}_u {\bf 8}_d{\bf {\overline {3}}}_s+ {\bf 3}_u {\bf 8}_d  {\bf {6}}_s +{\bf 3}_u {\bf 8}_d  {\bf {\overline {3}}}_s  \\
 & &+ {\bf {\overline 6}}_d{\bf 8}_u  {\bf {6}}_s +{\bf {\overline 6}}_d {\bf 8}_u{\bf {\overline {3}}}_s+ {\bf 3}_d {\bf 8}_u  {\bf {6}}_s+{\bf 3}_d {\bf 8}_u  {\bf {\overline {3}}}_s+ \cdots, \nonumber
 \end{eqnarray}
ignoring higher-dimensional terms involving four or more Higgs doublets.  
Each term comes with an overall coefficient, which we  leave implicit and  assume to be of order unity. The contraction of flavor indices is also implicit, {\it e.g.} the third term is $\epsilon^{ikl}{(\bf 3_u)}_i {(\bf 8_d)}^j_k  {({\bf {6}}_s)}_{jl} $.

For generic values of $\langle {\bf 6}_s \rangle$ and $\langle {\bf {\overline {3}}}_s \rangle$, this superpotential leaves only a single pair of Higgs doublets massless.  
Indeed, the ${\bf 6}_s$ multiplet alone is enough to do that job.  Writing ${\bf 8}_d$ as ${\bf 8}_d= \phi_a \lambda_a$, where $a$ takes the values 1 through 8, and the $\lambda_a$ are the generators of $SU(3)$, the $F$ component of $\phi_a$ is:
\begin{equation}
F_a = \frac{\partial W}{\partial \phi_a} = {\rm Tr}[{\bf {6}}_s( {\bf{\overline 6}}_u + A_u  )\lambda_a],
\end{equation}
where the antisymmetric $3 \times 3$ matrix $A_u$ is constructed according to $A_u^{ij}=\epsilon^{ijk}{{\bf 3}_u}_k$.
Unbroken supersymmetry at the flavor scale requires $F_a=0$ for all $a$ in the presence of the vacuum expectation value (VEV) of ${\bf {6}}_s$.   A short analysis shows that if ${\bf {6}}_s$ is invertible the solution has  ${\bf{\overline 6}}_u \propto {\bf {6}}_s^{-1} $ and $A_u =0$. 
As promised, there is only one flat direction.   Until we include condensation of ${\bf {\overline {3}}}_s$, however, the light up-type Higgs lives entirely in ${\bf{\overline 6}}_u$; similarly, the light down-type Higgs lives entirely in ${\bf{\overline 6}}_d$, indeed, ${\bf{\overline 6}}_d \propto {\bf {6}}_s^{-1} $ as well, and thus the up-type quark, down-type quark and lepton Yukawa coupling matrices are {\em all} proportional to one another. Taking into account both ${\bf 6}_s$ and ${\bf {\overline {3}}}_s$ condensations,  however, the two light Higgs fields will derive from linear combinations of ${\bf {\overline 6}}_u$ and  ${\bf 3}_u$, and  ${\bf {\overline 6}}_d$ and ${\bf 3}_d$, and realistic fermion masses and mixings can be achieved.  


Potentials for the flavor symmetry breaking fields can be generated through conventional supersymmetry-breaking effects.  In gravity mediated supersymmetry breaking \cite{Chamseddine:1982jx}, the effects of supersymmetry breaking are encoded through a spurion field $Z$ with an intermediate-scale $F$-component VEV, $\langle Z \rangle \sim v M_{pl} \theta^2$.  Potentials for the flavor-breaking fields arise from Kahler potential terms including
 \begin{eqnarray}
  {1 \over M_{Pl}^2}\int \! d^4 \theta Z^\dagger Z  {\bigg (}  {\rm Tr} [{\bf {6}}_s^\dagger {\bf {6}}_s]+{ {\rm Tr} [{\bf {6}}_s^\dagger {\bf {6}}_s]^2 \over M_{Pl}^2} & & \\
 +{{\rm Tr}[{\bf {6}}_s^\dagger {\bf {6}}_s{\bf {6}}_s^\dagger {\bf {6}}_s] \over M_{Pl}^2} + {\bf 3}_s^\dagger {\bf 3}_s+ {({\bf 3}_s^\dagger {\bf 3}_s)^2 \over M_{Pl}^2}   + \cdots {\bigg )}&&, \nonumber
\label{eq:spot}
\end{eqnarray}
where again we've suppressed numerical coefficients.  We assume that the potential is unstable at the origin so that the flavor symmetry is broken. With all couplings of order unity, one naively expects $f \sim M_{Pl}$, but we will assume that the flavor-breaking  VEVs are suppressed relative to the Planck scale roughly by an order of magnitude or two, for instance due to suppressed couplings for the terms quadratic in the flavor breaking fields. 
 (To  produce VEVs that lead to realistic fermion masses and mixings, it might be necessary to add extra flavor-breaking fields with  $U(1)_X$ charges that allow for additional potential terms.)  

To incorporate neutrino masses, we introduce right-handed neutrinos $N_i$ transforming as a triplet under $SU(3)$, with $U(1)_X$ charge 5.    Dirac neutrino masses arise from the obvious terms,
\begin{equation}
W =  \lambda_{\nu}^{(1)}  L_i {\bf {\overline 6}}_u^{ij} N_j+  \lambda_{\nu}^{(2)} \epsilon^{ijk}L_i { {{\bf { 3}}}_u}_{j}  N_k,
\label{eq:diracnumass}
\end{equation}
while Majorana masses for the right-handed neutrinos are generated by the higher-dimensional terms
\begin{eqnarray}
W & = & {1 \over M_{pl}}N_i N_j  [\epsilon^{ikm}  \epsilon^{jln}({\bf 6}_s)_{kl} ({\bf 6}_s)_{mn} \label{eq:nm} \\
&&+ {\bf {\overline {3}}}_s^i {\bf {\overline {3}}}_s^j + \epsilon^{ijk}  ({\bf 6}_s)_{kl}{\bf {\overline {3}}}_s^l]. \nonumber
\end{eqnarray}
These terms give right-handed neutrino masses of order $f^2/{M_{Pl}}$, supporting the usual  see-saw mechanism \cite{gellmann}.

At leading order, our $SU(3)$ flavor symmetry forces squark and slepton masses to be flavor universal.  Universality is spoiled at higher order, {\em e.g.}
\begin{equation}
{\mathcal L}_{soft} \supset {1 \over M_{pl}^2}\int \! d^4 \theta   Z^\dagger Z {Q^\dagger}^i Q_j \left(  \delta_i^j + { {{\bf 6}_s ^\dagger}^{jl}  {{{\bf 6}}_s}_{il}\over M_{Pl}^2}   + \cdots \right).
\end{equation} The extra terms lead to off-diagonal terms in the squark mass-squared matrix.  Depending on model-dependent details, experimental bounds on flavor-changing neutral currents might require small coefficients for these extra terms, but flavor symmetry has eased the problem.

It is clear from the superpotential of Eqn.~(\ref{eq:superpotential}) that any couplings involving the fermionic components of ${\bf 6}_s$ and ${\bf {\overline 3}}_s$  also involve either scalar or fermionic components of $\bf{8}_u$  or $\bf{8}_d$.  But all of these adjoint fields are very heavy, so there are no renormalizable couplings of the  fermion singlets to Higgs and Higgsino fields in the low-energy  theory.  By supersymmetry, the same follows for the  scalar components of ${\bf 6}_s$ and ${\bf {\overline 3}}_s$.  More complex models of this sort do support unsuppressed couplings, as we describe elsewhere \cite{ptw}; but now we turn to some different considerations, which suggest another possible source of such couplings. 


\section{$\mu$ term and Higgs couplings to flavor-breaking fields}
\label{sec:mu}

Flavor symmetry can forbid a bare $\mu$ term; indeed, our $SU(3)$ symmetry does.   This is a good starting-point, since the $\mu$ term must be ``unnaturally'' small in models with low-energy supersymmetry, but it is probably too much of a good thing, since detailed implementations of phenomenology appear to require a non-vanishing, roughly weak-scale $\mu$ term.  

To generate a $\mu$ term in our framework, we introduce a set of gauge-singlet flavor multiplets with $U(1)_X$ charge zero.  The $U(1)_X$ symmetry  guarantees that there will be no superpotential couplings
between these flavor multiplets and the ones whose condensations give mass to the heavy Higgs doublets.  On the other hand, because these flavor multipets are not charged under the global $U(1)_X$,
there is no reason why they can't have restoring potentials that stabilize their VEVs at weak-scale values rather than at large values of order $f$.   (Of course, the couplings of the Nambu-Goldstone bosons will still be suppressed by powers of $1/f$.) This makes them suitable for generating $\mu$. 
 
As a concrete example, consider $U(1)_X$-neutral multiplets ${\bf {\overline 6}}_s$ and ${\bf 3}_s$.  These fields couple to the light Higgses through the superpotential terms
\begin{equation}
W={\bf {\overline 6}}_s {\bf {\overline 6}}_u {\bf {\overline 6}}_d +  {\bf 3}_s {\bf {\overline 6}}_u {\bf 3}_d+ {\bf 3}_s {\bf {\overline 6}}_d {\bf 3}_u + {\bf 3}_s  {\bf 3}_u {\bf 3}_d + \cdots,
\label{eq:mu}
\end{equation}
and they also have couplings amongst themselves:
\begin{equation}
W={\bf {\overline 6}}_s {\bf {\overline 6}}_s {\bf {\overline 6}}_s+ {\bf 3}_s {\bf 3}_s {\bf {\overline 6}}_s + \cdots
\label{eq:restore}
\end{equation}
Suppose that the scalars in ${\bf {\overline 6}}_s$ receive positive weak-scale masses-squared from supersymmetry breaking, while the scalars in ${\bf 3}_s$ receive negative masses squared.
 Then the second term of Eqn.~(\ref{eq:restore}) generates  quartic terms in the potential for the components of ${\bf 3}_s$, which can stabilize the VEV of ${\bf 3}_s$ around the weak scale, and the interactions in Eqn.~(\ref{eq:mu}) generate a $\mu$ term of roughly the right size. 
 A contribution to the effective $B\mu$ term is generated by the square of the F-term for
 ${\bf {\overline 6}}_s$, $V \supset ({\bf 3}_s^*{\bf 3}_s^* {\bf {\overline 6}}_u {\bf {\overline 6}}_d + {\rm h.c.})$.  Additional contributions to the effective $B\mu$ term come from higher-dimensional terms such as
 \begin{equation}
 {1 \over M_{pl}}\int \! d^2 \theta    Z  \left( {\bf 3}_s  {\bf 3}_u {\bf 3}_d  +  {\bf 3}_s {\bf {\overline 6}}_u {\bf 3}_d+ {\bf 3}_s {\bf {\overline 6}}_d {\bf 3}_u  \right).
 \label{eq:Aterms}
 \end{equation}
 These interactions also generate weak-scale $A$ terms coupling the light singlets with the light Higgs doublets.  
 
One may worry that, because the potential derived from Eqns.~(\ref{eq:mu}, \ref{eq:restore}, \ref{eq:Aterms}) exhibits a $Z_3$ symmetry, formation of domain walls at temperatures around the electroweak scale will lead to cosmological problems.  However, this $Z_3$ could easily be spoiled by higher-dimensional operators.    For example, suppose that in addition to the flavor-breaking multiplets ${\bf 6}_s$ and ${\bf {\overline 3}}_s$ from the previous section, we also have ${\bf 3}_s'$, also with $U(1)_X$ charge -5.  Then nothing prevents operators like
\begin{equation}
\supset {1 \over M_{pl}^2}\int \! d^4 \theta   Z^\dagger Z \;{ {\bf 3}_s'}^\dagger {\bf 6}_s {\bf 3}_u  {\bf 3}_d + {\rm h.c.},
\end{equation}  which gives an additional contribution to the effective $B\mu$ term of order $v^2 f^2/M_{Pl}^2$, and explicity breaks the $Z_3$ symmetry.

A single linear combination, $S$, of all the light gauge-singlet superfields couples to $H_u$ and $H_d$, the Higgs doublets of the MSSM, determined as above, through an $S H_u H_d$ term.  The fermionic component of $S$ then mixes with the rest of the neutralinos due to electroweak symmetry breaking, just as in the next-to-minimal SSM (NMSSM) \cite{Nilles:1982dy}.  But the scalar component of $S$  will typically be a linear combination of non-degenerate mass eigenstates, and in addition to interaction
terms of the form $|S H_u|^2$ and $|S H_d|^2$ arising from the superpotential, there will also  be 
terms of the form $S' H_u H_d$ deriving from the couplings in Eqn.~(\ref{eq:Aterms}).  Through singlet and Higgs VEVs, both terms induce  mixings between the neutral Higgs bosons and the gauge-singlet scalars.  This mixing gives hope that the existence of one or more of the extra neutral scalars could be detected at the LHC, or at a future linear collider.  

\section{Incorporating a Peccei-Quinn Symmetry}
As might be expected from its motivation, the present framework easily accommodates an axionic solution to the strong-CP problem.  In the sort of model we present here, it is natural to expect that Peccei-Quinn symmetry is broken near the flavor-breaking scale, $f \sim 10^{16}$ GeV.  Conventionally, so large a breaking scale has been regarded as cosmologically suspect, but with inflation occurring after the Peccei-Quinn phase transition, anthropic arguments suggest it is not only viable, but even desirable \cite{Wilczek:2004cr}. 

To continue our $SU(3)$ example, we keep the same Higgs multiplets as and impose a global $U(1)_{PQ}$ symmetry under which these multiplets have charges  $\{ {\bf 3}_u(1),{\bf {\overline 6}}_u(1), {\bf 8}_u(-4) \}$ and $\{{\bf 3}_d(1),{\bf {\overline 6}}_d(1),{\bf 8}_d(-4) \}$. 
The precise charge of the flavor adjoint Higgs fields is not critical, although  to ensure that a pair of Higgs doublets remains light the charge should be different from that of the other Higgs multiplets.  
As before, all but one pair of light Higgs doublets becomes heavy from interactions with the flavor-breaking fields $\{ {\bf {6}}_s(3), {\bf {\overline {3}}}_s (3) \}$.    

To generate a $\mu$-term, we introduce gauge-singlet multiplets  $\{ {\bf {\overline 6}}_s(-2), {\bf 15}_s(4) \}$, with superpotential couplings
\begin{equation}
W= {\bf {\overline 6}}_s {\bf {\overline 6}}_u {\bf {\overline 6}}_d + {\bf {\overline 6}}_s {\bf {\overline 6}}_s  {\bf 15}_s + {\bf {\overline 6}}_s {\bf 3}_u {\bf 3}_d.
\label{eq:fifteen}
\end{equation}
Here the totally symmetric four-index tensor ${\bf 15}_s$ plays the role of the ${\bf {\overline 6}}_s$ from Sec.~\ref{sec:mu}, while the ${\bf {\overline 6}}_s$
takes the place of ${\bf 3}_s$.  We assume that a VEV for ${\bf {\overline 6}}_s$ is driven by a negative soft mass-squared, while ${\bf 15}_s$
has a positive soft mass-squared, allowing the second term of Eqn.~(\ref{eq:fifteen}) to stabilize the VEV of ${\bf {\overline 6}}_s$.  Note that their $U(1)_{PQ}$ charges prevent these multiplets from acquiring large masses from interactions with the flavor-symmetry breaking fields.  

The reason we use  different multiplets than those used to generate the $\mu$-term in the previous section becomes apparent when we
consider how the $U(1)_{PQ}$ breaking is transmitted to the Higgs sector.  For this purpose, we introduce another flavor-breaking multiplet ${\bf {\overline 3}}_s(2)$, which acquires a VEV of order $f$.  An effecive $B \mu$ term is then generated by
\begin{equation}
 {1 \over M_{pl}^3}\int \! d^4 \theta   Z^\dagger Z \;{\bf {\overline 3}}_s^\dagger \left( {\bf 3}_u  {\bf 3}_d +  {\bf {\overline 6}}_u {\bf 3}_d+ {\bf {\overline 6}}_d {\bf 3}_u \right) + {\rm h.c.}
\end{equation}

Note that the $SU(3)$ and  $U(1)_{PQ}$  symmetries allow the superpotential term
\begin{equation}
{1 \over M_{Pl}} ({\bf {\overline 3}}_s {\bf {\overline 6}}_s)^2,
\end{equation}
but this coupling gives masses of order $f$ only to three of the six components in ${\bf {\overline 6}}_s$.  The others are then free to acquire VEVs and generate a $\mu$-term in the way described above.  
If we had instead used ${\bf 3}_s$ to generate $\mu$ as in the previous section, the coupling ${\bf 3}_s {\bf {\overline 3}}_s$ would be allowed, and we would expect both multiplets to become heavy. 


\section{Comments}

The preceding discussions, though concrete example, suggest some synergistic connections between flavor symmetry and supersymmetry that may be much more general.   The existence of a small number of light Higgs doublets and of Peccei-Quinn symmetry, the approximate flavor universality of squark and slepton masses, and the emergence of appropriate $\mu$ and $B\mu$ terms are generally problematic issues that appear in new light from this perspective.    As negative virtues,  we easily avoid spoiling the unification of couplings or introducing dangerous Nambu-Goldstone fields.    And, we've argued, it is not implausible that some of the fields associated with flavor symmetry breaking are reasonably light and reasonably strongly coupled to light Higgs doublets, making them candidates for observation at future colliders.  

More specific and possibly realistic model-building along these lines, to incorporate gauge unification and desirable mass matrix textures, is well worthy of investigation.

\begin{acknowledgments} EB and DT-S thank Owen Simpson for useful conversations.  The work of EB and DT-S  was supported by a Cottrell College Science Award.
\end{acknowledgments}

\end{document}